\documentclass[twocolumn,journal]{revtex4-1}
\usepackage[LGR,T1]{fontenc}
\usepackage[latin9]{inputenc}
\usepackage{mathrsfs}
\usepackage{amsmath}
\usepackage{graphicx}
\usepackage[unicode=true,
 bookmarks=true,bookmarksnumbered=true,bookmarksopen=true,bookmarksopenlevel=1,
 breaklinks=false,pdfborder={0 0 1},backref=true,colorlinks=false]
 {hyperref}
\hypersetup{pdftitle={Your Title},
 pdfauthor={Your Name},
 pdfborderstyle=,pdfpagelayout=OneColumn,pdfnewwindow=true,pdfstartview=XYZ,plainpages=false,hypertex}
\begin{document}
\title{The Field Theory of Collective Cherenkov Radiation Associated with Electron Beams}
\author{Edl Schamiloglu\textsuperscript{1} and Alexander Figotin\textsuperscript{2} }
\affiliation{\textsuperscript{1}Department of Electrical and Computer Engineering,
University of New Mexico, Albuquerque, NM 87131-0001, USA~~~~\\
 \textsuperscript{2}Department of Mathematics, University of California,
Irvine, CA 92697, USA }
\begin{abstract}
Classical Cherenkov radiation is a celebrated physics phenomenon of electromagnetic (EM) radiation stimulated by an
electric charge moving with constant velocity in a three dimensional dielectric medium.  Cherenkov radiation has a wide 
spectrum and a particular distribution in space similar to the Mach cone created by a supersonic source.  It is also characterized
by the energy transfer from the charge's kinetic energy to the EM radiation.  In the case of an electron beam passing through the 
middle of a an EM waveguide, the radiation is manifested as collective Cherenkov radiation. In this case the electron beam
can be viewed as a one-dimensional non-neutral plasma whereas the waveguide can be viewed as a slow wave structure (SWS).
This collective radiation occurs in particular
in traveling wave tubes (TWTs), and it features the energy transfer from the electron beam to the EM radiation in the waveguide.  Based on a 
Lagrangian field theory, we develop a convincing argument that the collective Cherenkov effect in TWTs is, in fact,  a convective instability, that is,
amplification.  We also derive, for the first time, expressions identifying low- and high-frequency cutoffs for amplification in TWTs. 
\end{abstract}
%\keywords{TWT, Pierce theory, field theory, fundamental limitations on amplification.}

\maketitle
The collective Cherenkov effect is one of the fundamental mechanisms for stimulated emission of radiation
from electron beams propagating in media with slow waves \cite{Kuz1, Kuz2, Kuz3}, such as in a traveling wave tube (TWT) \cite{PierTWT}.  
It is well known that the mechanism of signal amplification in TWTs is based on the Cherenkov radiation effect occurring
in dielectric media. Though some features of Cherenkov radiation depend
on details of the dielectric environment there is one feature that
stands out as universal. This universal feature is manifested as a
higher speed of the electron flow compare to the characteristic velocity
in the dielectric medium. Using a Lagrangian field theory generalization
\citep{FigTWTbk, FR} of Pierce's TWT theory \citep{PierTWT} we establish
as its mathematical implication that the velocity of the electron
flow is always above the phase velocity of any TWT mode associated
with amplification. Remarkably, this statement holds for any conceivable
values of TWT parameters implying that the primary condition for Cherenkov
radiation is always fulfilled in our theory. The theory also yields, for the first time,
explicit formulas describing the low- and high-frequency cutoffs for
amplification. These cutoff frequencies depend on two significant
TWT parameters: (i) the ratio $\chi=\frac{w}{\mathring{v}}$ of the
phase velocity of the relevant mode of the slow wave structure (SWS)
$w$ and the velocity of the electron flow $\mathring{v}$; and (ii)
a single parameter $\gamma$ that integrates into it the intensity
of the electron flow and the strength of its interaction with the
SWS. It turns out that $\gamma=2\chi C_{\mathrm{P}}^{3}$ where $C_{\mathrm{P}}$
is the Pierce gain parameter \citep{PierTWT}. Interestingly, our
analysis shows that the commonly made assumption requiring the characteristic
velocity $w$ to be below the velocity of the electron flow $\mathring{v}$
is not necessary for amplification. In other words, even when $w$
is larger than $\mathring{v}$ the TWT modes associated with amplification
always have phase velocities that are below the electron flow velocity
$\mathring{v}$ in conformity with the primary Cherenkov radiation
requirement.

Important theoretical studies of the Cherenkov effect in TWTs conducted
in \citep{Kartash} resulted in the following significant conclusions:
(i) in the electrodynamics of plasmas and plasma-like media the collective
Cherenkov effect can be classified as related to wave--wave interactions
in which the energy of one of the interacting waves is negative; (ii)
the collective Cherenkov effect can be treated as one of numerous
electron beam instabilities; (iii) the fundamental role played by
plasma collective effects for Cherenkov radiation remains virtually
untouched in the case of the Cherenkov effect; (iv) the methods and
terminology of the general theory of instabilities developed in plasma
physics can be successfully applied to study the collective Cherenkov
effect. The TWT field theory and the results we obtained here are
consistent with these conclusions.

Our interest in Cherenkov radiation here is largely motivated by its
intimate relation with amplification in TWTs. This relation and its
importance was recognized by a number of researchers \citep{Kartash},
\citep{Gold}, \citep[Sec. 1.1-1.2, 4.4, 4.8-4.9, 7.3, 7.6; Chap. 8]{BenSweSch},
\citep[Sec. 8.2]{Tsimring}. The broadly treated phenomenon of Cherenkov
radiation is a fascinating subject with a rich history going back
to Heaviside and Sommerfeld \citep{Fermi}, \citep[Chap. 1, 2]{Afan},
\citep{Tamm58}, \citep{Ginzb}, \citep[Chap. 2]{Buts}. According
to Frank and Tamm \citep{FraTam}, Tamm \citep{Tamm39,Tamm58}, and
Ginzburg \citep{Ginzb}, the Vavilov-Cherenkov effect in a somewhat
narrow sense of the term is essentially radiation of electromagnetic
(EM) waves with a continuous spectrum and specific angular distribution
by an electric charge moving with constant velocity $v$ that exceeds
the phase speed of light $c^{\prime}\left(\omega\right)$ in the surrounding
medium under consideration, that is 
\begin{equation}
v>c^{\prime}\left(\omega\right)=\mathrm{c}/n\left(\omega\right),\label{eq:vcnom1a}
\end{equation}
where $n\left(\omega\right)$ is the index of refraction for light
at frequency $\omega$ in the medium and $\mathrm{c}$ is the velocity
of light in vacuum. The inequality (\ref{eq:vcnom1a}) plays an important
role in the Cherenkov effect and we name it the \emph{velocity inequality}.
This radiation is very directional and waves of a given frequency
$\omega$ are emitted at a specific angle $\varTheta$ to the direction
of motion of the system so that 
\begin{equation}
\cos\varTheta=\frac{c^{\prime}\left(\omega\right)}{v}.\label{eq:vcnom1b}
\end{equation}
Since for real-valued $\varTheta$ the value of $\left|\cos\varTheta\right|$
cannot exceed $1$, Eq. (\ref{eq:vcnom1b}) readily implies inequality
(\ref{eq:vcnom1a}). Or in other words, if the velocity inequality
does not hold, then (\ref{eq:vcnom1b}) cannot hold either for any
real $\varTheta$. The radiation has a clear-cut front which forms
a cone with the angle of opening $\pi-2\varTheta$ and the moving
charge at its apex. This cone is analogous to the Mach cone that characterizes
a shock wave generated by the motion of a supersonic source in air
or other media. The velocity of the shock wave or sound plays the
role of the phase velocity of light $c^{\prime}\left(\omega\right)$.
Note that the emitted EM waves must carry energy and, consequently,
energy conservation demands the particle kinetic energy reduction.
In other words the Cherenkov radiation assumes energy transfer
from the moving charge to EM waves.

The TWT is a vacuum electron device with a pencil-like electron beam
propagating on its axis. Therefore, it is quite natural to view the
TWT spatially as a one-dimensional (1D) continuum, and it is exactly
what Pierce did in his model \citep{PierTWT}, \citep[Sec. I]{Pier51}.
We generalized Pierce's theory as a 1D TWT field theory in \citep{FigTWTbk,FR}
adding to it space-charge effects. One of the goals we pursue here
is to find the relation between the Cherenkov effect and amplification
in TWTs as it is applicable to a 1D field theory. There are several
challenges in the pursuit of this goal. First, many well-established
features of Cherenkov radiation are special to three-dimensional (3D)
space, such as the Mach cone and the corresponding angle $\varTheta$
defined by Eq. (\ref{eq:vcnom1b}). Second, what exactly is the dielectric/polarizable
medium and $c^{\prime}\left(\omega\right)$ in the case of a 1D TWT
field theory? To answer this we need to expand the dielectric point
of view of the TWT and, in particular, we have to identify an analog
of the phase velocity of light $c^{\prime}\left(\omega\right)$. Third,
the inequality (\ref{eq:vcnom1a}), that is, $v>c^{\prime}\left(\omega\right)$,
can be viewed as a key property of the Cherenkov effect. In addition
to that, it also selects the frequencies for which the Cherenkov effect
can occur. The key question is whether this inequality is applicable
to a 1D theory provided the velocity $c^{\prime}\left(\omega\right)$
is identified, and if that is the case, what can we say about frequencies
for which it holds? Assume now that a satisfactory 1D version of the
Cherenkov effect is somehow constructed. We want to answer our main
question: what is the relation between the 1D version of the Cherenkov
effect and TWT amplification? In a nutshell, our approach to establishing
this relationship is as follows.

In all of our considerations frequency $\omega$ is assumed to be
real. Let us consider a TWT eigenmode $f_{\omega}\left(z,t\right)$
of frequency $\omega$ which is always of the form 
\begin{equation}
f_{\omega}\left(z,t\right)=a_{\omega}\exp\left\{ -\mathrm{i}\left(\omega t-kz\right)\right\} ,\quad k=k\left(\omega\right),\label{eq:fomtz1a}
\end{equation}
where $a_{\omega}$ is a complex-valued constant and $k\left(\omega\right)$
is a complex-valued wavenumber and its dependence on $\omega$ is
determined by the dispersion relation. If $k\left(\omega\right)$
is real then $f_{\omega}\left(z,t\right)$ is oscillatory, that is,
harmonic in space and time. In the case where the imaginary part of
$k$ is not equal to zero (($\Im\left(k\right)\neq0$), $f_{\omega}\left(z,t\right)$
exponentially grows or decays if $z\rightarrow\pm\infty$ and this
is a convective instability, in other words, amplification.

Let us now introduce a complex-valued phase velocity 
\begin{equation}
u=u\left(\omega\right)=\frac{\omega}{k\left(\omega\right)},\label{eq:fomtz1b}
\end{equation}
and note that since $\omega$ is real, the real part of $k$ ($\Re\left\{ k\right\} $)
is given by $\Re\left\{ k\right\} =\omega\Re\left\{ \frac{1}{u}\right\} $,
implying 
\begin{equation}
\Re\left\{ k\right\} =\frac{\omega}{\breve{\Re}\left\{ u\right\} },\quad\breve{\Re}\left\{ u\right\} \equiv\frac{\left|u\right|^{2}}{\Re\left\{ u\right\} }.\label{eq:fomtz1c}
\end{equation}
We call $\breve{\Re}\left\{ u\right\} $ defined by the second equality
in Eq. (\ref{eq:fomtz1c}) the ``pseudo-real part'' of the complex
number $u$ (see \citep[Chap. 57]{FigTWTbk}). It turns out that $\breve{\Re}\left\{ u\right\} $
is of physical significance, and can be identified with the wave's
energy velocity $u_{\mathrm{en}}$ \citep[Chap. 57]{FigTWTbk}. As
to the representation of the Cherenkov effect in the case of the TWT,
we make the following identifications: (i) $v$ is the velocity of
stationary electron flow in the TWT; and (ii) the phase velocity $c^{\prime}\left(\omega\right)$
is $\breve{\Re}\left\{ u\left(\omega\right)\right\} $ defined by
the second equation in Eq. (\ref{eq:fomtz1c}), that is, 
\begin{equation}
c^{\prime}\left(\omega\right)\equiv\breve{\Re}\left\{ u\left(\omega\right)\right\} .\label{eq:fomtz2a}
\end{equation}
To tie quantities $v$ and $\breve{\Re}\left\{ u\left(\omega\right)\right\} $
to features of the Cherenkov effect we proceed as follows. The common
way to express amplification in the TWT quantitatively is by relating
it to the so-called convective instability. The convective instability,
in turn, is manifested through exponential growth in space of TWT
eigenmodes $f_{\omega}\left(z,t\right)$ defined by Eq. (\ref{eq:fomtz1a}).
Then we can naturally quantify the rate of exponential growth of the
eigenmode $f_{\omega}\left(z,t\right)$ by $\Im\left(k\left(\omega\right)\right)$.
Consequently, $f_{\omega}\left(z,t\right)$ is a convective instability
mode if and only if 
\begin{gather}
\Im\left(k\left(\omega\right)\right)\neq0\text{ or equivalently,}\label{eq:fomtz2b}\\
\Im\left(u\left(\omega\right)\right)=-\omega\frac{\Im\left(k\left(\omega\right)\right)}{\left|k\left(\omega\right)\right|^{2}}\neq0,\nonumber 
\end{gather}
where $\Im\left(k\left(\omega\right)\right)$ is the imaginary part
of $k\left(\omega\right)$ and $\Im\left(u\left(\omega\right)\right)$
is the imaginary part of $u(\omega)$. Finally, our analysis shows
that the following mathematical implication always holds 
\begin{equation}
\Im\left(u\left(\omega\right)\right)\neq0\text{ implies }v>c^{\prime}\left(\omega\right),\label{eq:fomtz2c}
\end{equation}
or, in other words, if there is amplification then the velocity inequality
(\ref{eq:vcnom1a}) that is critical to the Cherenkov effect also
holds. Let us compare this fact with another implication related to
the Cherenkov effect, namely, Eq. (\ref{eq:vcnom1b}) implies the
velocity inequality (\ref{eq:vcnom1a}). This comparison suggests
that the relations in Eq. (\ref{eq:fomtz2b}) can be viewed as the TWT equivalent
of the critical Cherenkov effect (Eq. (\ref{eq:vcnom1b})). In other
words, we may view amplification as a representation of the Cherenkov
effect in a 1D field theory of a TWT. This mathematically precise
characterization of the Cherenkov effect in the 1D field theory of
a TWT can be strengthened further using the following physically transparent
argument. The electron beam in the TWT is a flow of electrons moving
with nearly constant velocity $v$. These electrons are moving close
to the SWS, which can be viewed as a polarizable medium similar to
a dielectric. This situation is known to lead to generation of EM
radiation accompanied by electrons losing kinetic energy, and it is
one of the well-known manifestations of the Cherenkov effect. The
synchronism and electron bunching boost the amplification, and, consequently
the Cherenkov effect.

The TWT field theory we use here is a generalization of the experimentally well-tested Pierce's theory
\citep{PierTWT}, \citep[Sec. I]{Pier51}. It was introduced and studied
in \citep[Chap. 4]{FigTWTbk}. We remind the reader that the celebrated
Pierce theory is a 1D theory of TWTs that accounts for the signal
amplification and the energy transfer from the electron flow to microwave
radiation \citep{Tsimring}, \citep[Chap. 4]{BBLN}, \citep[Chap. 4]{SchaB},
\citep{Gilm1}. Pierce's theory assumes: (i) an idealized linear representation
of the electron beam as a dynamic system; (ii) a lossless transmission
line (TL) representing the relevant eigenmode of the SWS that interacts
with the electron beam; (iii) the TL is spatially homogeneous with
uniformly distributed shunt capacitance and serial inductance. Further
modifications of Pierce's theory can be found in \citep{Sim17}.

The TWT field theory is a linear theory constructed based on Hamilton's principle
of least action. The states (configurations) of the electron beam
are conceived as perturbations of the stationary flow of electrons
represented by charge $q=q\left(z,t\right)$ with current $i\left(z,t\right)=\partial_{t}q\left(z,t\right)$.
The states of the TL are charge $Q=Q\left(z,t\right)$ with current
$I\left(z,t\right)=\partial_{t}Q\left(z,t\right)$. Then, the TWT
Lagrangian $\mathcal{L}{}_{\mathrm{TB}}$ is defined as follows \citep[Chaps. 4, 24]{FigTWTbk}:
\begin{gather}
\mathcal{L}{}_{\mathrm{TB}}=\mathcal{L}_{\mathrm{Tb}}+\mathcal{L}_{\mathrm{B}},\label{eq:LagTBqbe1a}\\
\mathcal{L}_{\mathrm{Tb}}=\frac{L}{2}\left(\partial_{t}Q\right)^{2}-\frac{1}{2C}\left(\partial_{z}Q+b\partial_{z}q\right)^{2},\nonumber \\
\mathcal{L}_{\mathrm{B}}=\frac{1}{2\beta}\left(\partial_{t}q+\mathring{v}\partial_{z}q\right)^{2}-\frac{2\pi}{\sigma_{\mathrm{B}}}q^{2},\nonumber 
\end{gather}
where $\mathcal{L}_{\mathrm{B}}$ represents the Lagrangian for the
electron beam and $\mathcal{L}_{\mathrm{Tb}}$ represents the Lagrangian
for the TL including the interaction with the electron beam. The parameter
$\beta$ in the electron beam Lagrangian $\mathcal{L}_{\mathrm{B}}$
is defined by 
\begin{equation}
\beta=\frac{\sigma_{\mathrm{B}}}{4\pi}R_{\mathrm{sc}}^{2}\omega_{\mathrm{p}}^{2}=\frac{e^{2}}{m}R_{\mathrm{sc}}^{2}\sigma_{\mathrm{B}}\mathring{n},\quad\omega_{\mathrm{p}}^{2}=\frac{4\pi\mathring{n}e^{2}}{m},\label{eq:LagTBqbe1b}
\end{equation}
where $-e$ is the electron charge with $e>0$, $m$ is the electron
mass, $\omega_{\mathrm{p}}$ is the electron beam plasma frequency,
$\sigma_{\mathrm{B}}$ is the area of the cross-section of the electron
beam, the constant $R_{\mathrm{sc}}$ is the so-called plasma frequency
reduction factor that accounts phenomenologically for the finite dimensions
of the electron beam cylinder as well as geometric features of the
SWS, and $\mathring{n}$ is the number density of electrons. Note
that the parameter $\mathring{v}$ in the electron beam Lagrangian
$\mathcal{L}_{\mathrm{B}}$ is the velocity of the stationary electron
flow, and the expression $\partial_{t}q+\mathring{v}\partial_{z}q$
is the so-called convective derivative to be expected for the ``Eulerian
point of view'' associated with the field theory. The term $-\frac{2\pi}{\sigma_{\mathrm{B}}}q^{2}$
in the Lagrangian $\mathcal{L}_{\mathrm{B}}$ represents space-charge
effects, particularly electron-to-electron repulsion \citep[Chaps. 4, 24]{FigTWTbk}

Parameters of the TL Lagrangian $\mathcal{L}_{\mathrm{Tb}}$ are:
(i) $C>0$ and $L>0$ are, respectively, its shunt capacitance and
inductance per unit length; (ii) $0<b\leq1$ is a phenomenological
parameter that couples the electron beam and the TL. Note that the
electron beam-TL interaction enters the Lagrangian $\mathcal{L}_{\mathrm{Tb}}$
through the term $\frac{1}{2C}\left(\partial_{z}Q+b\partial_{z}q\right)^{2}$
that has shunt capacitance $C$ as a factor associated physically
with the spatial gap between the electron beam surface and the wall
of the SWS. Consequently, the shunt charges $\partial_{z}Q$ and $\partial_{z}q$
enter the interaction term on equal footing except that the coupling
factor $b$ effectively reduces the inductive input of the electron
beam current, the coupling (see \citep[Chap. 3]{FigTWTbk} for more
details). Then, the corresponding Euler-Lagrange (EL) equations are
\begin{gather}
L\partial_{t}^{2}Q-\frac{1}{C}\partial_{z}^{2}\left(Q+bq\right)=0,\label{eq:LagTBqbe1c}\\
\frac{1}{\beta}\left(\partial_{t}+\mathring{v}\partial_{z}\right)^{2}q+\frac{4\pi}{\sigma_{\mathrm{B}}}q-\frac{b}{C}\partial_{z}^{2}\left(Q+bq\right)=0.\label{eq:LagTBbe1d}
\end{gather}
To analyze the EL equations we introduce TWT eigenmodes of the form
\begin{gather}
Q\left(z,t\right)=\hat{Q}\left(k,\omega\right)\mathrm{e}^{-\mathrm{i}\left(\omega t-kz\right)},\label{eq:LagTBbe2a}\\
q\left(z,t\right)=\hat{q}\left(k,\omega\right)\mathrm{e}^{-\mathrm{i}\left(\omega t-kz\right)},\nonumber 
\end{gather}
and apply to Eqs. (\ref{eq:LagTBqbe1c}) and (\ref{eq:LagTBbe1d})
the Fourier transform in time $t$ and for spatial variable $z$.
That and under the assumption that $\omega$ is fixed yields the following
eigenvalue problem
\begin{gather}
M_{u\omega}X=0,\quad X=\left[\begin{array}{r}
\hat{Q}\\
\hat{q}
\end{array}\right],\quad u=\frac{\omega}{k},\label{eq:LagTBbe2b}
\end{gather}
for a generalized eigenvalue $u$, which is the phase velocity, and
the corresponding eigenvector $X$, where $M_{u\omega}$ is a $2\times2$
matrix of the form 
\begin{equation}
M_{u\omega}=\left[\begin{array}{rr}
\frac{1}{u^{2}}-\frac{1}{w^{2}} & \frac{b}{u^{2}}\\
\frac{b}{u^{2}} & \:\left[\frac{1}{u^{2}}+\frac{1}{\gamma}\left(\frac{\omega_{\mathrm{rp}}^{2}}{\omega^{2}}-\frac{\left(u-\mathring{v}\right)^{2}}{u^{2}}\right)\right]b^{2}
\end{array}\right],\label{eq:LagTBbe2c}
\end{equation}
and the TWT principal parameter $\gamma$ is defined by 
\begin{equation}
\gamma=\frac{b^{2}}{C}\beta=\frac{b^{2}}{C}\frac{\sigma_{\mathrm{B}}}{4\pi}\omega_{\mathrm{rp}}^{2},\quad\omega_{\mathrm{rp}}=R_{\mathrm{sc}}\omega_{\mathrm{p}},\label{eq:LagTBbe2d}
\end{equation}
where $\omega_{\mathrm{rp}}$ is the so-called reduced plasma frequency.

The problem of finding the generalized eigenvalue $u$ is reduced
to the characteristic equation $\det\left\{ M_{u\omega}\right\} =0$,
which after elementary algebraic transformations turns into the following
characteristic equation 
\begin{gather}
\mathscr{D}\left(u,\gamma\right)=\frac{\gamma}{w^{2}-u^{2}}+\frac{\left(u-\mathring{v}\right)^{2}}{u^{2}}=\frac{1}{\breve{\omega}^{2}},\label{eq:LagTBbe3a}\\
\breve{\omega}=\frac{\omega}{\omega_{\mathrm{rp}}},\quad u=\frac{\omega}{k},\nonumber 
\end{gather}
and we refer to the function $\mathscr{D}\left(u,\gamma\right)$ as
the characteristic function. Note that the characteristic equation
in Eq. (\ref{eq:LagTBbe3a}) encodes all the information about the
dispersion relation in the form of a frequency-dependent phase velocity
$u\left(\omega\right)$. Indeed, if we find $u\left(\omega\right)$,
we can readily recover $k\left(\omega\right)=\frac{u\left(\omega\right)}{\omega}$,
which is the dispersion relation. We refer to solutions $u$ of the
characteristic equation Eq. (\ref{eq:LagTBbe3a}) as characteristic
velocities.

A convenient dimensionless form of the characteristic equation Eq.
(\ref{eq:LagTBbe3a}) is 
\begin{gather}
\mathscr{D}\left(\check{u},\check{\gamma}\right)=\frac{\check{\gamma}}{\chi^{2}-\check{u}{}^{2}}+\frac{\left(\check{u}-1\right)^{2}}{\check{u}{}^{2}}=\frac{1}{\breve{\omega}^{2}},\quad\breve{\omega}=\frac{\omega}{\omega_{\mathrm{rp}}},\label{eq:LagTBbe3b}\\
\check{\gamma}=\frac{\gamma}{\mathring{v}{}^{2}},\quad\check{u}=\frac{u}{\mathring{v}}=\frac{\omega}{k\mathring{v}},\quad\chi=\frac{w}{\mathring{v}}.\nonumber 
\end{gather}
Not to clutter notation we will use the same symbols for dimensionless
version of the parameters and variables as for most of the time the
dimensionless nature of equations is evident.

Pierce's theory emerges from our TWT field theory as its high-frequency
approximation, namely \citep[Chap. 4, 9, 62]{FigTWTbk} 
\begin{equation}
\mathscr{D}\left(u\right)=\frac{\check{\gamma}}{\chi^{2}-\check{u}{}^{2}}+\frac{\left(\check{u}-1\right)^{2}}{\check{u}{}^{2}}=0.\label{eq:LagTBbe3c}
\end{equation}
(It should be noted that this is the first time Pierce's theory has
been derived from first principles in a rigorous, self-consistent
manner.) The relation between our TWT principal parameter $\check{\gamma}$
and Pierce's gain parameter $C_{\mathrm{P}}$ is given by 
\begin{equation}
\check{\gamma}=\frac{\gamma}{\mathring{v}{}^{2}}=2\frac{w}{\mathring{v}}C_{\mathrm{P}}^{3}=2\chi C_{\mathrm{P}}^{3}.\label{eq:LagTBbe3d}
\end{equation}

Our TWT field theory reveals for the first time well-defined low-
and high-frequency cutoffs for amplification in TWTs. The TWT field
theory we present is constructed based on the principle of least action.
Therefore, energy conservation and energy transfer from the electron
beam to the EM radiation (represented by the state of the TL) are
exact and one may view the amplification frequency limits as fundamental.

An analysis of the characteristic equation Eq. (\ref{eq:LagTBbe3c})
shows that when $\chi<1$ there exists a critical value $\gamma_{\mathrm{Pcr}}>0$
of the parameter $\chi$ such that 
\begin{enumerate}
\item for $0<\gamma<\gamma_{\mathrm{Pcr}}$ and all solutions $u$ to Eq.
(\ref{eq:LagTBbe3c}) are real-valued and there is no amplification; 
\item for $\gamma>\gamma_{\mathrm{Pcr}}$ there are exactly two different
real-valued solutions $u$ to Eq. (\ref{eq:LagTBbe3c}) and exactly
two different complex-valued solutions that are complex-conjugate
so that there is amplification. 
\end{enumerate}
Consequently, for the case when $\chi<1$ amplification is possible
if and only if $\gamma>\gamma_{\mathrm{Pcr}}$ and, if that is the
case, it occurs for all frequencies.

\begin{figure}[h]
\begin{centering}
\includegraphics[scale=0.35]{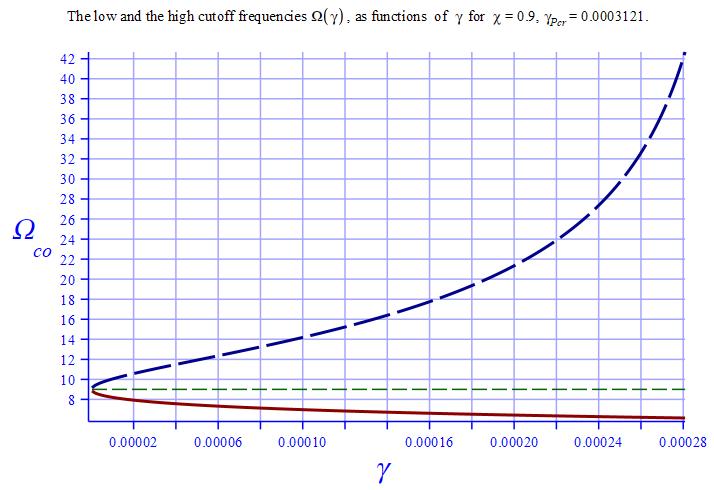} 
\par\end{centering}
\caption{\label{fig:lowup-coff}Plot of the low-frequency cutoff $\Omega_{-}\left(\gamma,\chi\right)$
(solid red curve) and the high-frequency cutoff $\Omega_{+}\left(\gamma,\chi\right)$
(dashed blue curve) for $\chi=\frac{w}{\mathring{v}}=0.9<1$ and for
$0.1\gamma_{\mathrm{Pcr}}\left(0.9\right)<\gamma<0.9\gamma_{\mathrm{Pcr}}\left(0.9\right)$.
Note that high the frequency $\Omega_{+}\left(\gamma,\chi\right)$
approaches infinity as $\gamma$ approaches the critical value $\gamma_{\mathrm{Pcr}}\left(\chi\right)$,
that is $\lim_{\gamma\rightarrow\gamma_{\mathrm{Pcr}}\left(\chi\right)}\Omega_{+}\left(\gamma,\chi\right)=+\infty$.
The horizontal dash (green) straight line identifies the $\lim_{\gamma\rightarrow0}\Omega_{-}\left(\gamma,\chi\right)=\lim_{\gamma\rightarrow0}\Omega_{+}\left(\gamma,\chi\right)=\frac{\chi}{1-\chi}=9$.
Hence, the plot illustrates, in particular, the inequalities in Eq.
(\ref{eq:OmchiOm1a}) and the limit relations in Eq. (\ref{eq:OmchiOm1b}).}
\end{figure}

The critical value $\gamma_{\mathrm{Pcr}}$ and its corresponding
critical value $u_{\mathrm{Pcr}}$ are intimately related to Pierce's
theory and they satisfy the relations 
\begin{gather}
u_{\mathrm{Pcr}}\left(\chi\right)=\chi^{\frac{2}{3}},\quad\gamma_{\mathrm{Pcr}}\left(\chi\right)=\left.\frac{\left(u^{2}-{\it \chi}^{2}\right)\left(u-1\right)^{2}}{u^{2}}\right|_{u=u_{\mathrm{Pcr}}}\label{eq:charP1b-1}\\
=\left(1-\chi^{\frac{2}{3}}\right)^{3}=\left(1-u_{\mathrm{Pcr}}\left(\chi\right)\right)^{3}.\nonumber 
\end{gather}
For the case when $\chi=\frac{w}{\mathring{v}}<1$ there exist two
fundamental phase velocities $u_{\mp}\left(\gamma,\chi\right)$ which
are stationary points of the characteristic function $\mathscr{D}\left(u\right)$
satisfying the relations 
\begin{equation}
\partial_{u}\mathscr{D}\left(u\right)=0,\quad\mathscr{D}\left(u\right)>0,\quad u=u_{\mp}\left(\gamma,\chi\right),\label{eq:duDin1a}
\end{equation}
\begin{equation}
0<u_{-}\left(\gamma,\chi\right)<\chi;\quad\chi<u_{+}\left(\gamma,\chi\right)<1;\label{eq:duDin1b}
\end{equation}
that is, $u_{\mp}$ are extremum points and $\mathscr{D}\left(u_{\mp}\right)$
are the corresponding extreme values of the characteristic function
$\mathscr{D}\left(u\right)$. Then, having $u_{\mp}$ and Eq. (\ref{eq:duDin1a})
in mind, we introduce the following two frequencies 
\begin{equation}
\Omega_{\mp}\left(\gamma,\chi\right)=\frac{1}{\sqrt{\mathscr{D}\left(u_{\mp}\left(\gamma,\chi\right)\right)}}.\label{eq:duDin1c}
\end{equation}
We refer to frequencies $\Omega_{-}\left(\gamma,\chi\right)$ and
$\Omega_{+}\left(\gamma,\chi\right)$ as the low- and high-frequency
cutoffs for the instability, respectively (see Figure \ref{fig:lowup-coff}).
The names are justified by the fact that, if $u$ is a characteristic
velocity, then the following implication holds 
\[
\Im\left\{ u\right\} \neq0\Leftrightarrow\Omega_{-}\left(\gamma,\chi\right)<\omega<\Omega_{+}\left(\gamma,\chi\right).
\]

\begin{figure*}
\begin{centering}
\hspace{0.1cm}\includegraphics[scale=0.3]{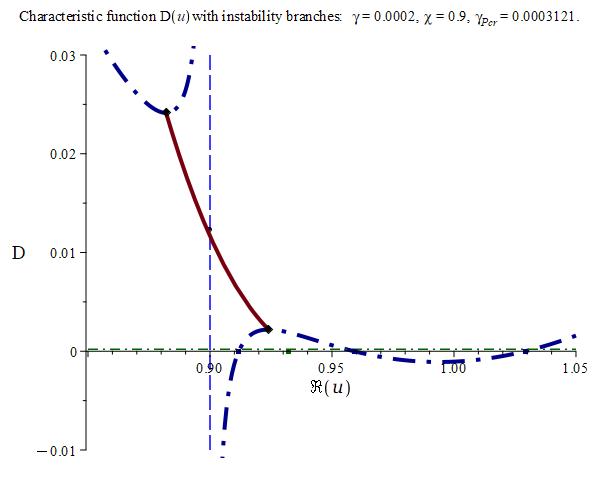}\hspace{0.1cm}\hspace{2cm}\hspace{0.1cm}\includegraphics[scale=0.3]{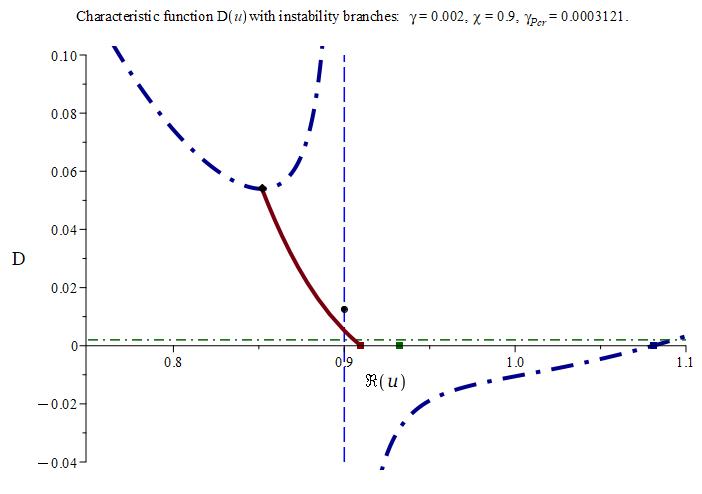}\hspace{0.1cm} 
\par\end{centering}
\centering{}(a)\hspace{7cm}(b)\caption{\label{fig:char-Du-frag} Plots of fragments of the characteristic
function $\mathscr{D}\left(u\right)$ with integrated instability
branches for $\chi=0.9$ and $\gamma=0.002>\gamma_{\mathrm{Pcr}}\protect\cong0.0003121$
(horizontal axis -- $\Re\left\{ u\right\} $, vertical axis -- $D$):
(a) for $\gamma=0.0002<\gamma_{\mathrm{Pcr}}\protect\cong0.0003121$;
(b) for $\gamma=0.002>\gamma_{\mathrm{Pcr}}\protect\cong0.0003121$.
Solid (brown) curves represent instability branches with $\Im\left\{ u\right\} \protect\neq0$,
dash-dotted (blue) curves represent oscillatory branches with $\Im\left\{ u\right\} =0$,
vertical dashed straight lines represent the asymptotes of $\mathscr{D}\left(u\right)$.
The instability nodes (points of transition from stability to instability)
are represented by solid (brown) square dots.}
\end{figure*}

Figure \ref{fig:char-Du-frag} shows the fragments of the characteristic
function $\mathscr{D}\left(u\right)$ with the extremum points.
Figure \ref{fig:char-Du-frag}(a) shows the case when $\gamma=0.0002<\gamma_{\mathrm{Pcr}}\cong0.0003121$
when both cutoff frequencies $\Omega_{\mp}\left(\gamma,\chi\right)$
are finite whereas in the case when $\gamma=0.002>\gamma_{\mathrm{Pcr}}\cong0.0003121$
we have $\Omega_{+}\left(\gamma,\chi\right)=+\infty$ with the corresponding
$\mathscr{D}\left(u_{+}\left(\gamma,\chi\right)\right)=0$ as one
can see in Figure \ref{fig:char-Du-frag}(b). 

\begin{figure*}
\begin{centering}
\hspace{-0.5cm}\includegraphics[scale=0.4]{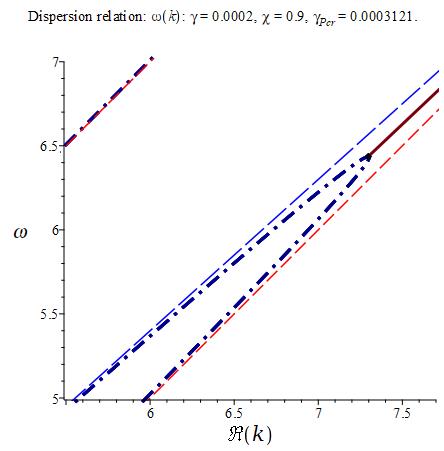}\hspace{1cm}\includegraphics[scale=0.4]{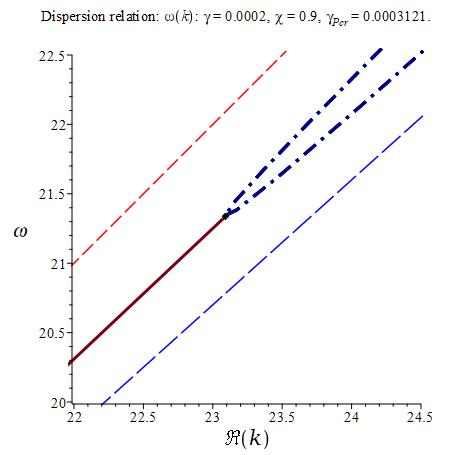} 
\par\end{centering}
\centering{}(a)\hspace{6.5cm}(b)\caption{\label{fig:disp} Fragments of the dispersion-instability graph (vertical
axis -- $\omega$, horizontal axis -- $\Re\left\{ k\right\} $)
for $\chi=0.9$, $\gamma=0.0002<\gamma_{\mathrm{Pcr}}\protect\cong0.0003121$:
(a) for positive phase velocities about the low frequency cutoff;
(b) for positive phase velocities about the high frequency cutoff.
Solid (brown) curves represent unstable branches with $\Im\left\{ k\right\} \protect\neq0$,
dash-dotted (blue) curves represent oscillatory branches with $\Im\left\{ k\right\} =0$,
dashed straight lines represent the dispersion relations of the cold
TL (blue) and cold electron beam (red). The instability nodes (points
of transition from stability to instability) are represented by solid
(brown) boxes. The frequencies of the instability nodes for $\Re\left\{ k\right\} >0$
are the low- (a) and high-frequency (b) cutoffs.}
\end{figure*}

The dimensionless characteristic equation (\ref{eq:LagTBbe3b}) allows
to completely analyze the set of non-real characteristic velocities
associated with the TWT instability and amplification. The low- and
the high-frequency cutoffs' behavior as a function of $\gamma$ depends
significantly on whether $\chi<1$ or $\chi>1$. A summary of our
analysis of the low- and the high-frequency cutoffs $\Omega_{-}\left(\gamma,\chi\right)$
and $\Omega_{+}\left(\gamma,\chi\right)$ is as follows. For the case
when $\chi=\frac{w}{\mathring{v}}<1$, the cutoff frequencies $\Omega_{-}\left(\gamma,\chi\right)$
and $\Omega_{+}\left(\gamma,\chi\right)$ satisfy the following inequalities:
\begin{equation}
\Omega_{-}\left(\gamma,\chi\right)<\frac{\chi}{1-\chi}<\Omega_{+}\left(\gamma,\chi\right),\quad\gamma>0,\quad\chi<1,\label{eq:OmchiOm1a}
\end{equation}
and the following limit identities: 
\begin{equation}
\lim_{\gamma\rightarrow0}\Omega_{-}\left(\gamma,\chi\right)=\lim_{\gamma\rightarrow0}\Omega_{+}\left(\gamma,\chi\right)=\frac{\chi}{1-\chi},\quad\chi<1,\label{eq:OmchiOm1b}
\end{equation}
\begin{equation}
\lim_{\gamma\rightarrow\gamma_{\mathrm{Pcr}}\left(\chi\right)}\Omega_{+}\left(\gamma,\chi\right)=+\infty,\quad\lim_{\gamma\rightarrow\infty}\Omega_{-}\left(\gamma,\chi\right)=0,\quad\chi<1.\label{eq:OmchiOm1c}
\end{equation}
For the case $\chi=\frac{w}{\mathring{v}}>1$ the high-frequency cutoff
$\Omega_{+}\left(\gamma,\chi\right)$ is infinite for any $\gamma$,
that is 
\begin{equation}
\Omega_{+}\left(\gamma,\chi\right)=+\infty,\quad\chi>1,\label{eq:OmchuOm1d}
\end{equation}
and the following limit relations hold for the low-frequency cutoff
$\Omega_{-}\left(\gamma,\chi\right)$: 
\begin{equation}
\lim_{\gamma\rightarrow0}\Omega_{-}\left(\gamma,\chi\right)=\infty,\quad\lim_{\gamma\rightarrow\infty}\Omega_{-}\left(\gamma,\chi\right)=0,\quad\chi>1.\label{eq:OmchiOm1e}
\end{equation}
Figure \ref{fig:lowup-coff} shows the low- and high-frequency cutoffs
for $\chi=\frac{w}{\mathring{v}}=0.9$ illustrating, in particular,
inequalities (\ref{eq:OmchiOm1b}) and the limit relations (\ref{eq:OmchiOm1b})
by means of the dashed (green) horizontal straight line. Figure \ref{fig:disp}
shows the fragments of the dispersion-instability graph (see the definition
below) indicating the low- and the high-frequency cutoffs $\Omega_{-}\left(\gamma,\chi\right)$
and $\Omega_{+}\left(\gamma,\chi\right)$, respectively. 

It is useful to integrate the information about the TWT instability
into the dispersion relations using the concept of a dispersion-instability
graph that we developed in \citep[Chap. 7]{FigTWTbk}. Recall that
conventional dispersion relations are defined as the relations between
the real-valued frequency $\omega$ and the real-valued wavenumber
$k$ associated with the relevant eigenmodes. In the case of the convective
instability, frequency $\omega$ is real and wavenumber $k$ is complex-valued.
To represent the corresponding modes geometrically as points in the
real $\omega-k$ plane we proceed as follows. In this case we parametrize
every mode of the TWT system uniquely by the pair $\left(k\left(\omega\right),\omega\right)$.
In view of the importance to us of the mode instability, that is,
when $\Im\left\{ k\left(\omega\right)\right\} \neq0$, we partition
all the system modes represented by pairs $\left(\omega,k\left(\omega\right)\right)$
into two distinct classes -- oscillatory modes and unstable ones
-- based on whether the wavenumber $k\left(\omega\right)$ is real-
or complex-valued with $\Im\left\{ k\left(\omega\right)\right\} \neq0$.
We refer to a mode (eigenmode) of the system as an oscillatory mode
if its wavenumber $k\left(\omega\right)$ is real-valued. We associate
with such an oscillatory mode point $\left(\omega,k\left(\omega\right)\right)$
in the $\omega-k$ plane with $\omega$ being the vertical axis and
$k$ being the horizontal one. Similarly, we refer to a mode (eigenmode)
of the system as a convectively unstable mode if its wavenumber $k=k\left(\omega\right)$
is complex-valued with a nonzero imaginary part, that is, $\Im\left\{ k\left(\omega\right)\right\} \neq0$.
We associate with such an unstable mode point $\left(\omega,(\Re\left\{ k\left(\omega\right)\right\} \right)$
in the $\omega-k$ plane.

Based on the above considerations, we represent the set of all oscillatory
and convectively unstable modes of the system geometrically by the
set of the corresponding modal points $\left(\omega,k\left(\omega\right)\right)$
and $\left(\omega,\Re\left\{ k\left(\omega\right)\right\} \right)$
in the $\omega-k$ plane. We name this set the dispersion-instability
graph. To distinguish graphically points $\left(\omega,k\left(\omega\right)\right)$
associated with oscillatory modes when $k\left(\omega\right)$ is
real-valued from points $\left(\omega,\Re\left\{ k\left(\omega\right)\right\} \right)$
associated with unstable modes when $k\left(\omega\right)$ is complex-valued
with $\Im\left\{ k\left(\omega\right)\right\} \neq0$, we show points
$\Im\left\{ k\left(\omega\right)\right\} =0$ in blue color whereas
points with $\Im\left\{ k\left(\omega\right)\right\} \neq0$ are shown
in brown color. We remind once again that every point $\left(\omega,\Re\left\{ k\left(\omega\right)\right\} \right)$
with $\Im\left\{ k\left(\omega\right)\right\} \neq0$ represents exactly
two complex conjugate convectively unstable modes associated with
$\pm\Im\left\{ k\left(\omega\right)\right\} $. 

Finally, the low- and high-frequency cutoffs that we have identified
for amplification in TWTs have recently been verified in particle-in-cell
(PIC) simulations \citep{Rouh23}. Plans for an experimental campaign
to validate the theoretical and PIC simulation results are under way
\citep{Nus1, Nus2}.

In conclusion, we present results from a Lagrangian field theory generalization
of Pierce's TWT theory that convincingly shows that the Cherenkov
effect in TWTs is a convective instability leading to amplification.
We derive expressions for the first time that identify low- and high-frequency
cutoffs for amplification in TWTs. These results can be tested in experiment and will prove valuable
in designing future TWT experiments and in explaining experimental
observations where TWT amplifiers transition from amplification to
oscillation, as we will describe in our next publication \citep{SchFig}.

\emph{Appendix.} In this Appendix, we provide a precise mathematical
argument proving that the velocity of any TWT eigenmode associated
with amplification is strictly below the velocity of the electron
flow which is unity in dimensionless units. Suppose $u$ is a complex-valued
characteristic velocity satisfying the characteristic equation Eq.
(\ref{eq:LagTBbe3b}) and relations 
\begin{equation}
\Re\left\{ u\right\} >0,\quad\Im\left\{ u\right\} \neq0.\label{eq:comuR1a}
\end{equation}
We claim then that $u$ also satisfies the inequalities 
\begin{equation}
0<\Re\left\{ u\right\} <\breve{\Re}\left\{ u\right\} \equiv\frac{\left|u\right|^{2}}{\Re\left\{ u\right\} }<1.\label{eq:comuR1b}
\end{equation}
We call $\breve{\Re}\left\{ u\right\} $ in Eq. (\ref{eq:comuR1b})
the ``pseudo-real part'' of the complex number $u$ (see \citep[Chap. 57]{FigTWTbk}).
This turns out to be of physical significance, and can be identified
with the wave's energy velocity $u_{\mathrm{en}}$ \citep[Chap. 57]{FigTWTbk}.
Note first that inequality $\Re\left\{ u\right\} <\breve{\Re}\left\{ u\right\} $
always holds as long as $\Re\left\{ u\right\} >0$. It remains to
show that $\breve{\Re}\left\{ u\right\} <1$. Note also that $\Im\left\{ \mathscr{D}\left(\check{u},\check{\gamma}\right)\right\} =0$
since the right-hand-side $\frac{1}{\breve{\omega}^{2}}$ of the characteristic
Eq. (\ref{eq:LagTBbe3b}) is real-valued. After tedious but elementary
algebraic transformations, Eq. $\Im\left\{ \mathscr{D}\left(\check{u},\check{\gamma}\right)\right\} =0$
can be transformed into 
\begin{gather}
\left(1-\breve{\Re}\left\{ u\right\} \right)\left[\left(\Re\left\{ u\right\} +\chi\right)^{2}+\left(\Im\left\{ u\right\} \right)^{2}\right]\times\label{eq:comuR1c}\\
\left[\left(\Re\left\{ u\right\} -\chi\right)^{2}+\left(\Im\left\{ u\right\} \right)^{2}\right]=\check{\gamma}\left|u\right|^{4}.\nonumber 
\end{gather}
Since $\check{\gamma}$ and all terms in Eq. (\ref{eq:comuR1c}) except
for $1-\breve{\Re}\left\{ u\right\} $ are {\emph{a priori}} positive,
the desired inequality $\breve{\Re}\left\{ u\right\} <1$ must hold
and that completes the argument.

The physical significance of inequalities (\ref{eq:comuR1b}) is that
they assure that if (i) the characteristic velocity $u$ corresponds
to an unstable eigenmode and, consequently, $\Im\left\{ u\right\} \neq0$
and (ii) its real phase velocity $\Re\left\{ u\right\} $ is positive,
then $\Re\left\{ u\right\} <\breve{\Re}\left\{ u\right\} <1$, manifesting
that the eigenmode velocity is always below the velocity of the electron
flow which is unity in dimensionless units.

This research was supported by AFOSR MURI under Grant No. FA9550-20-1-0409
administered through the University of New Mexico.

\end{document}